\documentclass[useAMS,usenatbib,preprint2]{aastex}
\usepackage{amsmath,amssymb,wasysym}
\usepackage{delarray}
\usepackage{natbib}
\usepackage{graphicx}
\usepackage{url}
\newcommand{\fesc}{\rm f_{esc}}
\newcommand{\xion}{\rm x_{ion}}
\newcommand{\zreion}{\rm z_{reion}}

\newcommand{\mdm}{m_{\rm dm}}
\newcommand{\hmpc}{$\rm{h^{-1}Mpc}$ }

\newcommand{\xm}{\langle \rm{x} \rangle^{\rm{m}}}
\newcommand{\xv}{\langle \rm{x} \rangle^{\rm{v}}}
\newcommand{\hmo}{{\rm h^{-1}}}

\newcommand{\Msun}{\ensuremath{\mathrm{M}_{\odot}}}

\newcommand{\nicefrac}[2]{\leavevmode\kern.1em
            \raise.5ex\hbox{\the\scriptfont0 #1}\kern-.1em
      /\kern-.15em\lower.25ex\hbox{\the\scriptfont0 #2}}

\begin{document}

%\title{The stratified reionization history of the Milky Way halo: radiative tranfer in a progenitor of our Galaxy}
%\shorttitle{Reionization of a Milky Way progenitor}
%\title{The internal and isolated reionization of local group galaxies}
%\title{The isolated and internal reionization of a Milky Way progenitor: CLUES/ATON simulations}
\title{High resolution simulations of the reionization of an isolated Milky Way - M31 galaxy pair}
\shorttitle{ATON meets CLUES}

%\author[P.\ Ocvirk]{P. Ocvirk$^{{1}}$\\
\author{P. Ocvirk\altaffilmark{1}, D. Aubert\altaffilmark{1}, J. Chardin\altaffilmark{1}, A. Knebe\altaffilmark{2}, N. Libeskind\altaffilmark{3}, S. Gottl\"{o}ber\altaffilmark{3}, G. Yepes\altaffilmark{2} and Y. Hoffman\altaffilmark{4}}
\affil{
$^1$ Observatoire Astronomique de Strasbourg, 11 rue de l'Universit\'e, 67000 Strasbourg, France}
\affil{
$^2$ Grupo de Astrof\'{i}sica, Departamento de Fisica Teorica, Modulo C-8, Universidad Aut\'{o}noma de Madrid, Cantoblanco E-280049, Spain}
\affil{
$^3$ Leibniz-Institute f\"{u}r Astrophysik Potsdam (AIP), An der Sternwarte 16, D-14482 Potsdam, Germany}
\affil{
$^4$ Racah Institute of Physics, Hebrew University, Jerusalem 91904, Israel}

%\date{Typeset \today; Received / Accepted}

%\pagerange{\pageref{firstpage}--\pageref{lastpage}} \pubyear{2007}

%\maketitle
%\label{firstpage}

%\onecolumn

\begin{abstract}
We present the results of a set of numerical simulations aimed at studying reionization at galactic scale. We use a high resolution simulation of the formation of the Milky Way-M31 system to simulate the reionization of the local group.
The reionization calculation was performed with the post-processing radiative transfer code ATON and the underlying cosmological simulation was performed as part of the CLUES project\thanks{{http://www.clues-project.org}}. We vary the source models to bracket the range of source properties used in the literature. We investigate the structure and propagation of the galatic ionization fronts by a visual examination of our reionization maps.
Within the progenitors we find that reionization is patchy, and proceeds locally inside out. The process becomes patchier with decreasing source photon output. It is generally dominated by one major HII region and 1-4 additional isolated smaller bubbles, which eventually overlap. 
Higher emissivity results in faster and earlier local reionization. In all models, the reionization of the Milky Way and M31 are similar in duration, i.e. between 203 Myr and 22 Myr depending on the source model, placing their $\zreion$ between 8.4 and 13.7.
%Suppressing star formation in haloes less massive than $10^{9} {\rm \Msun}$ leads to a delayed reionization with respect to our baseline models, with less structure, fewer patches within the progenitors.
%The internal versus external nature of the reionization of the Milky Way and M31 progenitors is strongly model-dependent: when considering all atomic cooling haloes as sources, the 2 galaxies reionize in isolation for all emissivities, despite their proximity all along cosmic history. On the contrary, in a scenario where supernovae would provide a strong internal feedback, leaving only massive haloes (${\rm M}>10^{9} {\rm \Msun}$) to be efficient sources, we find that the Milky Way can be reionized by M31.
In all models except the most extreme, the MW and M31 progenitors reionize internally, ignoring each other, despite being relatively close to each other even during the epoch of reionization.
%The corresponding reionization maps always show a clear gap in $\zreion$ between the two progenitors. 
Only in the case of strong supernova feedback suppressing star formation in haloes less massive than $10^{9} {\rm \Msun}$, and using our highest emissivity, we find that the MW is reionized by M31.
\end{abstract}


%
%________________________________________________________________

%\setcounter{page}{1}

%\tableofcontents

%\vfill

%%%%%%%%%%%%%%%%%%%%%%%%%%%%%%%%%%%%%%%%%
\section{Introduction}
%%%%%%%%%%%%%%%%%%%%%%%%%%%%%%%%%%%%%%%%%
In the last decade, the epoch of reionization (hereafter EoR) has received increasing attention. Most observational works now seem to converge on reionization beginning as early as z=15 \citep{kogut2003short} and finishing around z=6 \citep{fan2006}, in apparent agreement with theoretical expectations \citep{haardt2011}. Reionization also affects the way galaxies form: it has been suggested that the rising metagalactic UV radiation field is responsible for evaporating the gas of low-mass galaxies \citep{gnedin2000,hoeft2006}, affecting their star formation and therefore also the buildup of the galactic halo \citep{bekkichiba2005}. This seems to provide a credible solution to the ``missing satellites problem'' \citep{klypin1999,moore1999}, by inhibiting star formation in low mass galaxies at early times \citep{bullock2000,benson2002a,benson2002b,benson2003}.
In this framework, a number of simple semi-analytical models (hereafter SAMs) have been shown to reproduce well the satellite population of the Milky Way (hereafter MW), such as \cite{koposov2009,munoz2009,busha2010,maccio2010,li2010,font2011}. 
They suggest that the  ultra-faint dwarf galaxies (hereafter UFDs) discovered by the SDSS \citep{martin2004,willman2005short,zucker2006,belokurov2007short, irwin2007short, walsh2007} are effectively reionization fossils, living in sub-haloes of about $10^{6-9} \Msun$.
More recently, \cite{ocvirk2011,ocvirk2012} showed that the structure of the UV background during reionization has a strong impact on the properties of the satellite population of galaxies. In particular, they showed that an internally-driven reionization led to significant changes in the radial distribution of satellites. 
%However, their simple semi-analytical model did not allow to account for an accurate description of the complex propagation of multiple ionization fronts. 
It is therefore of prime importance to determine how inhomogeneous the UV field is within a MW progenitor during reionization in a realistic setting. 
It has been shown that at least at large scales, reionization is a highly patchy process \citep{zahn2007,aubert2010}.
% and indicate that for a MW-mass galaxy it could take place at almost any time between 
%, as shows for instance Fig. 3 of  \citep{zahn2007} or Fig. 1 of \citep{lunnan2011}. 
%However, one must note that our Galaxy typically lives within one of the patches of these figures. Moreover, each of the patches naturally grows inside-out. 
%However, large scale simulations do not resolve all the UV sources within these patches, and their sphericity and inside-out growth could therefore be exaggerated in these works. 
However, little is known on how reionization proceeds at galactic scales. For instance, is it driven by internal or external sources?
%Indeed, individual galaxy-sized patches in large scale simulations grow roughly inside-out, suggesting  the existence of gradients of reionization redshift within a given galaxy.
%ut large scale simulations do not resolve all the UV sources within these patches, and their sphericity and inside-out growth could therefore be exaggerated in these works. 
%What is the effect of the lighter haloes on the geometry of reionization at galaxy scale? Do these additional sources wash out the radial reionization gradient? Is it preserved due to the clustering of these sources near the progenitor center? Do MW-size galaxies reionize due to internal or external sources?
Works from \cite{weinmann2007,alvarez2009} and \cite{iliev2011} (hereafter I11) attempted to describe reionization at galactic scale, but focused on rather large volumes, in order to account for large, rare overdensities such as galaxy clusters and groups, which produce the earliest sources. In these studies, the spatial resolution of the radiative transfer (hereafter RT) grid was 0.25-1 \hmpc, which does not allow a detailed study of the ionization fronts (hereinfater I-front) propagation {\em within} a MW progenitor. Moreover, the Milky Way environment also comprises the nearby, massive Andromeda galaxy. Therefore, despite these previous studies, a number of questions are left open, including:
\begin{itemize}
\item{What is the influence of our neighbour M31 on the reionization of our Galaxy?}
\item{What is the structure of the UV field at galactic scales? Approximately isotropic and/or symmetric, or very inhomogeneous?}
%\item{Could it be that the distribution of the sources would lead to a quasi-uniform UV field?}
%\item{Is reionization at galaxy scale monotonous and continuous or chaotic, with recombination events?}
\end{itemize}
Here we address these questions by focusing on a small volume containing the progenitors of the major local group (hereafter LG) galaxies (MW, M31, M33). We simulate its reionization with a 21 $h^{-1}$kpc spatial resolution, to gain insight into the development and overlap of I-fronts inside the volume of the MW progenitor and its direct neighbourhood.
The paper is laid out as follows: first we describe the simulation used and radiative postprocessing technique (Sec. \ref{s:methodology}). We then proceed to our results (Sec. \ref{s:results}), and discuss them (Sec. \ref{s:discussion}), before presenting our conclusions.

\section{Methodology}
%%%%%%%%%%%%%%%%%%%%%%%%%%%%%%%%%%%%%%%%%
\label{s:methodology}

\begin{table*}
\begin{center}
 \begin{tabular}{cccccccccc} 
 \hline 
 \hline 
 (1) & (2) & (3) & (4) & \multicolumn{2}{c}{(5)} &  \multicolumn{2}{c}{(6)} & \multicolumn{2}{c}{(7)} \\ 
 Model & Source & Source & Emissivity & \multicolumn{2}{c}{$\zreion^m$} & \multicolumn{2}{c}{$\Delta {\rm z}_{0.1}^{0.9}$} & \multicolumn{2}{c}{$\Delta {\rm t}$ (Myr)}\\ 
 name & type & criterion & (photons/s/$h^{-1}\Msun$) & MW & M31 & MW & M31 & MW & M31\\ 
 
 \hline 
 \hline 
 SPH & SPH star & - & $ 6.3 \times 10^{45}$ & 9 & 9.4 & 2.72 & 2.34 & 202 & 159 \\ 
 \hline
 H42 & DM halo & ${\rm T_{vir}>10^4 K}$ & $ 6.8 \times 10^{42}$ & 8.4 & 8.4 & 2.42 & 2.45 & 203 & 203 \\ 
 H43 & DM halo & ${\rm T_{vir}>10^4 K}$ & $ 6.8 \times 10^{43}$ & 11.5 & 11.8 & 2.57 & 2.48 & 105 & 98 \\ 
 H44 & DM halo & ${\rm T_{vir}>10^4 K}$ & $ 4.08 \times 10^{44}$ & 13.6 & 13.7 & 3.2 & 1.8 & 89 & 54 \\ 
 \hline
 H42 SNfb & DM halo & ${\rm M>10^9 h^{-1}\Msun}$ & $ 6.8 \times 10^{42}$ & 5.8 & 5.8 & 1.47 & 1.49 & 304 & 299 \\ 
 H43 SNfb & DM halo & ${\rm M>10^9 h^{-1}\Msun}$ & $ 6.8 \times 10^{43}$ & 8.3 & 9.3 & 1.5 & 1.37 & 138 & 109 \\ 
 H44 SNfb & DM halo & ${\rm M>10^9 h^{-1}\Msun}$ & $ 4.08 \times 10^{44}$ & 9.1 & 9.7 & 0.55 & 0.32 & 43 & 22 \\ 
 \hline 
 \end{tabular}

\caption{Properties of the models used. Note that the emissivity is given in photons/s/$\Msun$ of mass of {\em young stars} for the SPH model, and of DM halo for the halo-based models. It is given {\em after} accounting for an escape fraction $\fesc=0.2$. The H42 model has been calibrated so as to produce the same total number of photons as the SPH model at z=10. Column (4) gives the mass threshold of the DM halo based source models, used to mimic various forms of feedback. Column (5) gives the reionization redshift of the MW and M31 progenitors for each model, i.e. the time when the mass-weighted ionized fraction of the progenitor shown in Fig. \ref{f:xs} reaches 0.5. Column (6) gives the duration of the progenitors' reionization as the time spent to increase the mass-weighted ionized fraction $\xm$ from 0.1 to 0.9. Column (7) gives this duration in Myr.}
%To compute it we consider the MW progenitor is inscribed in a 1 \hmpc radius sphere centered on the center of mass. The MW reionization redshift is the time when half of the progenitor cells are reionized.}
\label{t:models}
\end{center}
\end{table*}

In this paper we use a radiative postprocessing method. A N-body-SPH hydrodynamical simulation of cosmic structure formation provides the gas distribution and ionizing sources distribution. These are the inputs to the radiative transfer code ATON \citep{aubert2008}, which computes the propagation of the photons and the evolution of the ionized fraction of the gas.

\subsection{The CLUES simulation}
\label{s:simclues}
The simulation used in this work was performed in the framework of the CLUES project \citep{clues2010}\footnote{Simulation seed number 186592}. 
It was run using standard Lambda cold dark matter ($\Lambda$CDM) initial conditions assuming a WMAP3 cosmology \citep{spergel2007}, i.e. $\Omega_{\rm{m}}=0.24$, $\Omega_{\rm{b}}=0.042$, $\Omega_{\Lambda}=0.76$. A power spectrum with a normalization of $\sigma_8=0.73$ and $n=0.95$ slope was used. The PMTree-SPH MPI code GADGET2 \citep{springel2005} was used to simulate the evolution of a cosmological box with a side length of $64$ \hmpc. Within this box a model Local Group that closely resembles the real Local Group was identified using a 1024$^3$ particles run (cf. \cite{libeskind2010}). This Local Group was then re-sampled with 64 times higher mass resolution in a region of 2 \hmpc about its center giving an equivalent resolution of 4096$^3$ particles, i.e. a mass resolution of $\mdm=2.1 \times 10^5 h^{-1} \Msun$ for the dark matter and $m_{gas}=4.42 \times 10^4 h^{-1} \Msun$ for the gas particles. For more detail we refer the reader to \cite{clues2010}. The feedback and star formation prescriptions of \cite{springel2003} were used. Outputs are written on average every 30 Myr.
The simulation starts at z=50. As it runs, dark matter and gas collapse into sheets and filaments, extending between halos, as comprehensively described in \cite{ocvirk2008,codis2012,hoffman2012,libeskind2012}. These feed proto-galaxies which then start forming stars.
It includes a uniform rising UV cosmic background generated from quasi-stellar objects and active galactic nuclei \cite{haardtmadau96}, switched on at z=6. Therefore the radiative transfer computations that we perform will be valid only at earlier times. We will see that this is not a problem, since our models always achieve complete reionization before z=6.
This simulation has been used to investigate a number of properties of galaxy formation at high resolution \citep{jaime2011,knebe2011b,knebe2011a,libeskind2011a,libeskind2011b}. Besides being a well-studied simulation, the advantage of this dataset for the present study is twofold. First of all, it produces a fairly realistic local group at z=0: the MW and M31 are in the correct range of masses and separations. Secondly, its mass resolution in the zoomed region allows us to resolve the $10^7 h^{-1}\Msun$ haloes. This is of crucial importance in reionization studies since they are the most numerous sources of UV photons.

\subsection{Radiative post-processing}
\subsubsection{ATON}
ATON is a post-processing code that relies on a moment-based description of
the radiative transfer equations and tracks the out-of-equilibrium ionisations
and cooling processes involving atomic hydrogen \citep{aubert2008}. Radiative quantities (energy
density, flux and pressure) are described on a fixed grid and evolved
according to an explicit scheme under the constraint of a Courant-Friedrich-Lewy condition (hereafter CFL). The
simulations presented in this work used a mono-frequency treatment of the
radiation with a typical frequency of 20.27 eV for 50000 K black body
spectrum. Because of the high resolution of the CLUES simulation, we do not make any correction for the clumping, as was done for the largest boxes of \cite{aubert2010}. ATON has been ported on multi-GPU architecture, where each GPU
handles a cartesian sub-domain and communications are dealt using the MPI
protocol \citep{aubert2010}. By achieving an x80 acceleration factor compared to CPUs, the CFL
condition is satisfied at high resolution within short wallclock computing
times. As a consequence, no reduced speed of light approximation is necessary
and it may be of great importance for the timing arguments of the local
reionization discussed hereafter.
Along the course of this work, simulations were run on segments of 8 to
64 GPUs on the Titane and Curie machines of the CCRT/CEA supercomputing facility, with typically 160000 radiative timesteps performed in 37 hours.

The postprocessing approximation has potentially important consequences on our results, as discussed for instance in \cite{baek2009,frank2012}. While the temperature of the gas is consistently followed by ATON, the gas density is ``frozen'' to that given by the SPH simulation snapshots. 
This means that our scheme does not allow for photo-evaporation, but the photo-heating does affect the radiative transfer calculations, e.g. through the temperature dependence of the recombination rates. In this sense, our scheme is "thermally coupled", which is already an important step in the direction of a more realistic treatment \citep{iliev2006c,pawlik2011}.
By design, self-shielding is also accounted for, and results in a later reionization of sourceless high density regions, such as mini-haloes or the cold gas filaments. However, in a fully coupled radiative-hydrodynamics simulation, the gas field reacts to the photo-heating, and can result in the dispersion of low-mass gas structures \citep{shapiro2004,iliev2005,iliev2009}. This should induce a form of self-regulation of star formation and therefore emissivity by shutting off sources in the ionized low-mass haloes, as shown in \cite{iliev2007}. Even though a small number of coupled galaxy formation codes have recently been built \citep{petkova2011,rosdahl2011,finlator2011,enzomoray}, at the moment no application to the formation of the local group in a zoom simulation such as the CLUES dataset we use here has been performed, mainly because of the huge computational cost involved. In any case, the impact of full coupling on our results is very likely negligible compared to uncertainties in the source efficiency, which prompts us to explore a wide range of emissivities (see Sec. \ref{s:sources}).

\subsubsection{Fields setup}
The gas density field is projected onto a $512^3$ grid of 11 \hmpc side.
The center of the grid is the barycenter of all the particles which end up within 300 kpc of the MW at z=0. 
%This allows us to define and follow the center of the MW progenitor.
This setup gives us a spatial resolution of ${\Delta x} = 21 \, {\rm{ h^{-1} kpc}}$. The sources are projected on the same grid. 
As explained in Sec. \ref{s:simclues}, the CLUES simulation uses a zoom technique, with a high and low resolution domains. The high resolution (hereafter HR) domain contains the objects of interest (MW and M31), and is described with dark matter, gas and star particles. At $512^3$ resolution all grid cells contain at least one gas particle in the HR region in the highest redshift snapshot (z=19.5). On the other hand, the low resolution (hereafter LR) domain does not have any SPH particle. Therefore we set the gas density in the low resolution domain to $\rho_{\rm LR}= 10^{-2} \rho_{\rm C}$, where $\rho_{\rm C}$ is the critical density of the Universe. The LR region does not contain any stars either. Photons reaching the HR/LR boundary region just  leave the local group and quickly reach the edges of the computational box. There, we use transmissive boundary conditions, i.e. light just exits the box.

\begin{figure}[t]
  {\includegraphics[width=1.\linewidth,clip]{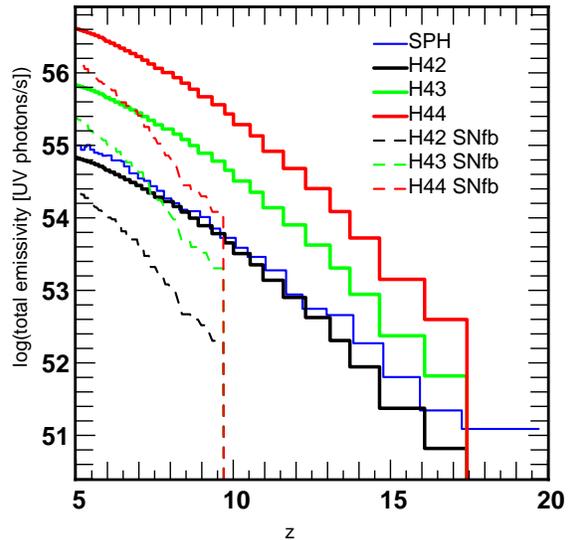}}
\caption{Total emissivities of the sources in the 11 \hmpc radiative transfer box for our 7 models.}
\label{f:totphot}
\end{figure}

\subsubsection{Ionizing source models}
\label{s:sources}
We use different source models based on either the star particles spawned by the hydrodynamical star formation prescriptions of the CLUES simulation, or dark matter haloes. 
%In both cases we are in the low efficiency cases of I11, and we have to stay there.
%We assume that a rate SFR=1\Msun/yr produces $1.6 \, 10^{53}$ ionising photons/s as in \cite{shapiro2004}.
We use a constant $\fesc=0.2$, which is among values allowed by recent studies on the UV continuum escape fraction of high-z galaxies \citep{wise2009,razoumov2010,yajima2011}. We neglect any possible AGN-phase of our emitters. Such sources could already be in place in rare massive proto-clusters during reionization \citep{dubois2011,dubois2012}, and contribute to the cosmic budget of ionizing photons \citep{haardt2011}, but they are beyond the scope of the present study.
% We defer the study of their influence during EoR to a later paper. 
The properties of our source models are summarized in Tab. \ref{t:models}.

\paragraph{SPH stars:}
\label{s:SPH}
In this model our sources are the star particles formed in the CLUES simulation.
At each time step we select all star particles younger than 60 Myr.
We use the ionizing luminosity of Tab. 2 of \cite{baek2009}, simulation S20. 
While the massive stars responsible for this radiation would have a typical lifetime of 10 Myr, we dilute their emission over 60 Myr in order to reduce the stochastic nature of star particle spawning in the simulation. Each star particle weighs $m_{\star}=2.21 \times 10^4 h^{-1} \Msun$. It is therefore more representative of a ``cluster'' of stars, and results in 1.9$\times 10^{50}$ ionizing photons/s delivered to the IGM for 60 Myr, after accounting for $\fesc=0.2$.

Because of the large mass of the star particles, and the slow pace of star formation during the EoR, we can not guarantee that the star formation rate is converged in the numerical sense \citep{springel03,rasera2006}, and as a consequence some sites of early star formation may be missing. Moreover, \cite{chardin2012} showed that the topology of HII regions during the EoR could be strongly affected by this issue.
Therefore we also consider source models based on dark matter haloes.

%The disadvantage is stochasticity, but the advantage is the SFR within a given galaxy includes self-consisten regulation by SN feedback for instance. 

\paragraph{Halo sources:}
\label{s:iliev}

As an alternative to the direct use of SPH stars, we here use dark matter haloes as sources. They are detected using the Amiga halo finder \citep{AHF04,AHF}. We keep only the haloes which have more than 90\% of their mass in high resolution  dark matter particles. Dwarf galaxies of the early Universe  are subject to a wide range of feedback processes beyond photo-evaporation by a UV background. Although our code does not allow for live self-regulation of the sources, we tried to account for the influence of at least some of the relevant feedback processes.

\subparagraph{External Lyman-Werner background\\}

Massive stars radiate in the Lyman-Werner (LW) band, which, contrary to H-ionizing photons, can travel several tens of Mpc through neutral hydrogen \citep{barkana2001,ahn2009}. Therefore it is legitimate to consider that our computational domain containing the LG must see the LW radiation of the earliest, rare sources of the Universe. It has been shown that the stellar LW background can be sufficiently strong to dissociate molecular hydrogen, which is the main coolant of pristine low mass haloes below ${\rm T_{vir}=10^4 K}$, therefore suppressing star formation. This can happen very early during the EoR, as early as z=12 according to \cite{ahn2009,ahn2012}. However, the latter is an average value, and it is only natural to expect that the LW background rises faster in overdensities such as the LG, all the more so in the neighborhood of a massive cluster such as Virgo. Therefore, we consider that only the atomic cooling haloes in our box are able to form stars. Following \cite{iliev2002}, we consider that haloes need to have a virial temperature ${\rm T_{vir}>10^4 K}$ in order to do so. We use their formula to compute the minimum mass required for a halo to be an atomic cooling source. It gives roughly $1.4 \times 10^7 \hmo \Msun$ at z=20 and about $1 \times 10^8 \hmo \Msun$ at z=5, which is the time spanned by our radiative post-processing simulation. In this model, only the haloes above this mass are emitting.
We assign an instantaneous star formation rate to each halo, assuming SFR $\propto  M$. 
Note that in our framework, unlike in I11, sources are not regulated, i.e. above the mass threshold, they emit continuously as long as the halo exists.
In order to be able to compare the SPH and halo-based models we calibrated the emissivity of the haloes of the H42 model so that the total number of UV photons emitted at z=10 is the same in both models. With this calibration, the total photon output evolution of the 2 models is quite similar, although SPH is slightly more luminous, as can be seen in Fig. \ref{f:totphot}.
%Note however that the computational volume contains a handful of massive galaxies which do not belong to our LG, and therefore the total photon output shown is representative of this ``extended'' LG, not just the MW-M31-M33.
%At high z, we have many more halo sources than SPH stars formed. For instance, at z=18.5, we have 9 SPH sources but 26 halos in the local group progenitor. 
%With our calibration, the average luminosity of the halo sources is lower: the most luminous halo is roughly 3 times fainter than in the SPH source model.
Accounting for a constant $\fesc=0.2$, our calibration gives for H42:
\begin{equation}
%{\rm \frac{\dot{N}_{\gamma}}{M}= 6. 10^{41} {\rm photons/s/\Msun}}
{\rm \dot{N}_{\gamma} / M}= 6.8 \times 10^{42} {\rm photons/s/\Msun}
\end{equation}
The emissivities of the other models are listed in Tab. \ref{t:models}, and their total photon output are shown in Fig. \ref{f:totphot} as well. 
%In contrast to e.g. I11, who uses different efficiencies for low and high mass haloes, we use the same source efficiency throughout the simulation and for all haloes. 
In our units, the low (high) efficiency assumptions of I11 give about $4-6 \times 10^{42-43}$ ($4 \times 10^{43}  - 1 \times 10^{44}$) photons/s/$\Msun$. Therefore the range of emissivities explored here roughly brackets the scenarios considered in I11 and extend to the lower efficiency scenario of \cite{baek2009}. These have been shown to produce viable reionization histories in their respective contexts, and are therefore used here as guidance, although the more abundant much smaller halos.

\paragraph{Strong feedback from supernovae\\}
Stellar feedback processes in galaxies are expected to affect star formation, but are generally poorly constrained. Supernovae, for instance, by blowing away and/or heating their host galaxy's cold gas reservoir, have been shown to be potentially very important for the evolution of low mass galaxies \citep{benson2003}. The details of supernovae feedback and the galaxy mass at which it kicks in, depend on a wide range of parameters (stellar rotation and chemical composition determining supernova energy production, mechanical coupling to the host's gas, host dark matter density profile, gas metallicity and its cooling properties) and has been hotly debated since \citep{dekel1986}. At low redshift, some insight can be gleaned thanks to direct observation of the interaction between supernova ejecta or superbubbles and their host's gas. Such observations are not available at high redshift, let alone during the EoR. Quantifying their impact on the host galaxy is therefore all the more difficult. Furthermore, supernova feedback at high z could be quite different from what we see at low z \cite{barkana2001}.
Recent semi-analytical models of galaxy formation \citep{kim2013} during the EoR have shown that supernovae feedback could render star formation in low-mass galaxies (${\rm V_{circ} < 30 }$ km/s) so inefficient that their contribution to the ionizing photon budget becomes negligble (less than 1\%). At the beginning of our radiative post-processing, (i.e. z=19.35), this corresponds to dark matter haloes of roughly $10^9 \hmo {\rm \Msun}$. 
%Using a semi-analytical model accounting for a strong supernova feedback, \cite{raivcevic2011} showed that the major contributors to the ionizing photon budget are the $10^9 {\rm \Msun}$ haloes.
In order to investigate the impact of such strong supernovae feedback we designed an additional set of models, which we refer to as HXX SNfb. Following the results of \cite{kim2013}, all haloes less massive than $10^9 \hmo {\rm \Msun}$ are assumed to be inefficiently forming stars and we set their emissivity to 0.
Above this mass threshold, we assume that galaxies are allowed to continuously form stars and produce UV photons following the same emissivities as H42,43,44.\footnote{To prevent any confusion, we here remind the reader that this strong feedback process operates exclusively in the radiative post-processing step of our workflow. The original SPH simulation remains the same as in the baseline H42-44 and SPH models.}
Because of the small number of emitters, the total photon output of this model within our box is much smaller than in the other 4 baseline scenarios, as can be seen in Fig. \ref{f:totphot}. The figure shows that, in this model, efficient sources appear only at z$<10$, whereas the SPH model, which also includes a self-consistent supernova feedback, forms stars at all times. This readily shows our proposed implementation of supernova feedback in the HXX SNfb models is rather extreme, which is why we refer to it as ``strong'' feedback. Nevertheless, we consider it as a limiting case.
% a model of the strongest suppressive feedback supernovae could provide.

\begin{figure}
%\begin{tabular}{c}
{\includegraphics[width=0.95\linewidth,clip]{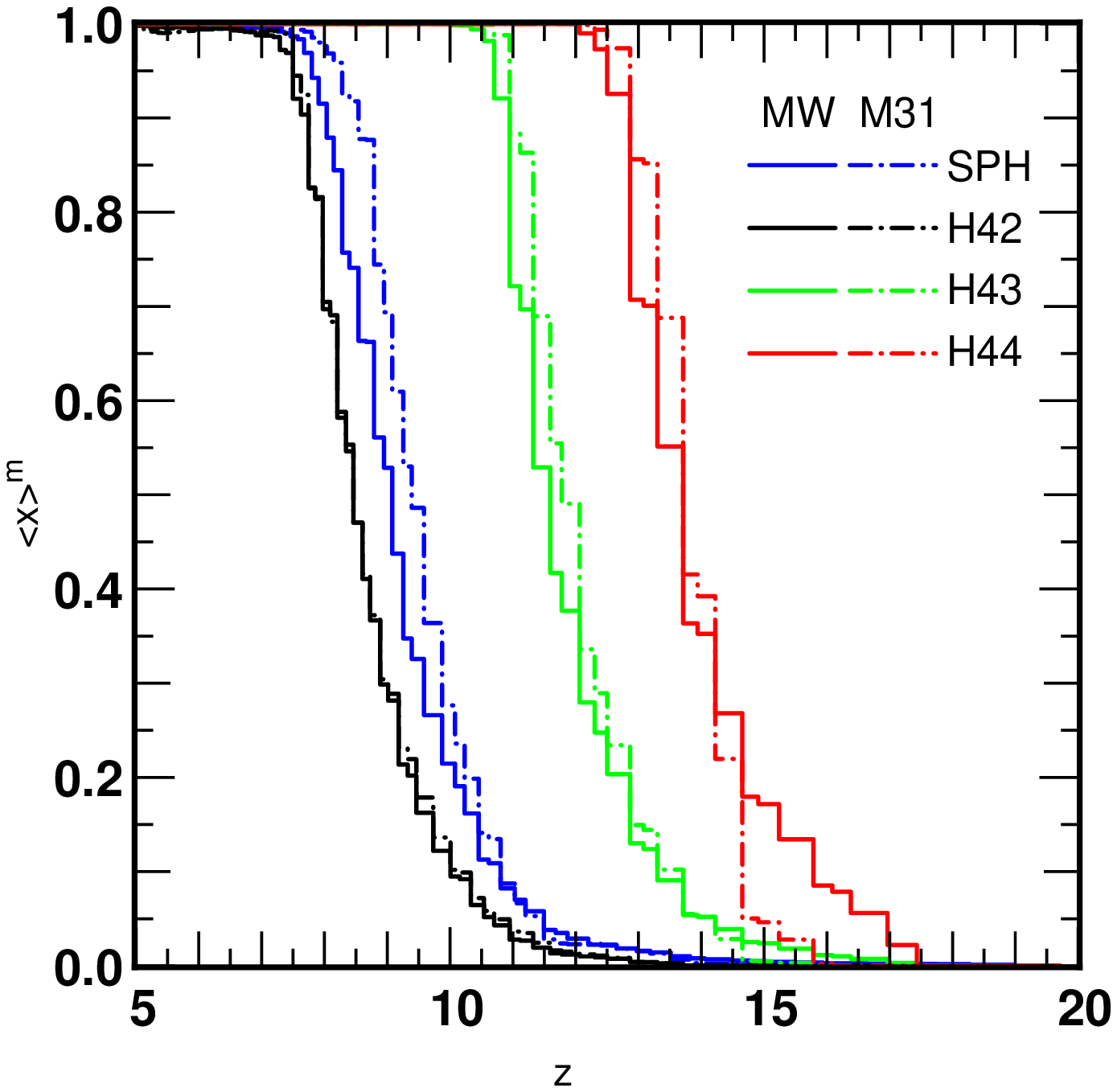}}\\
{\includegraphics[width=0.95\linewidth,clip]{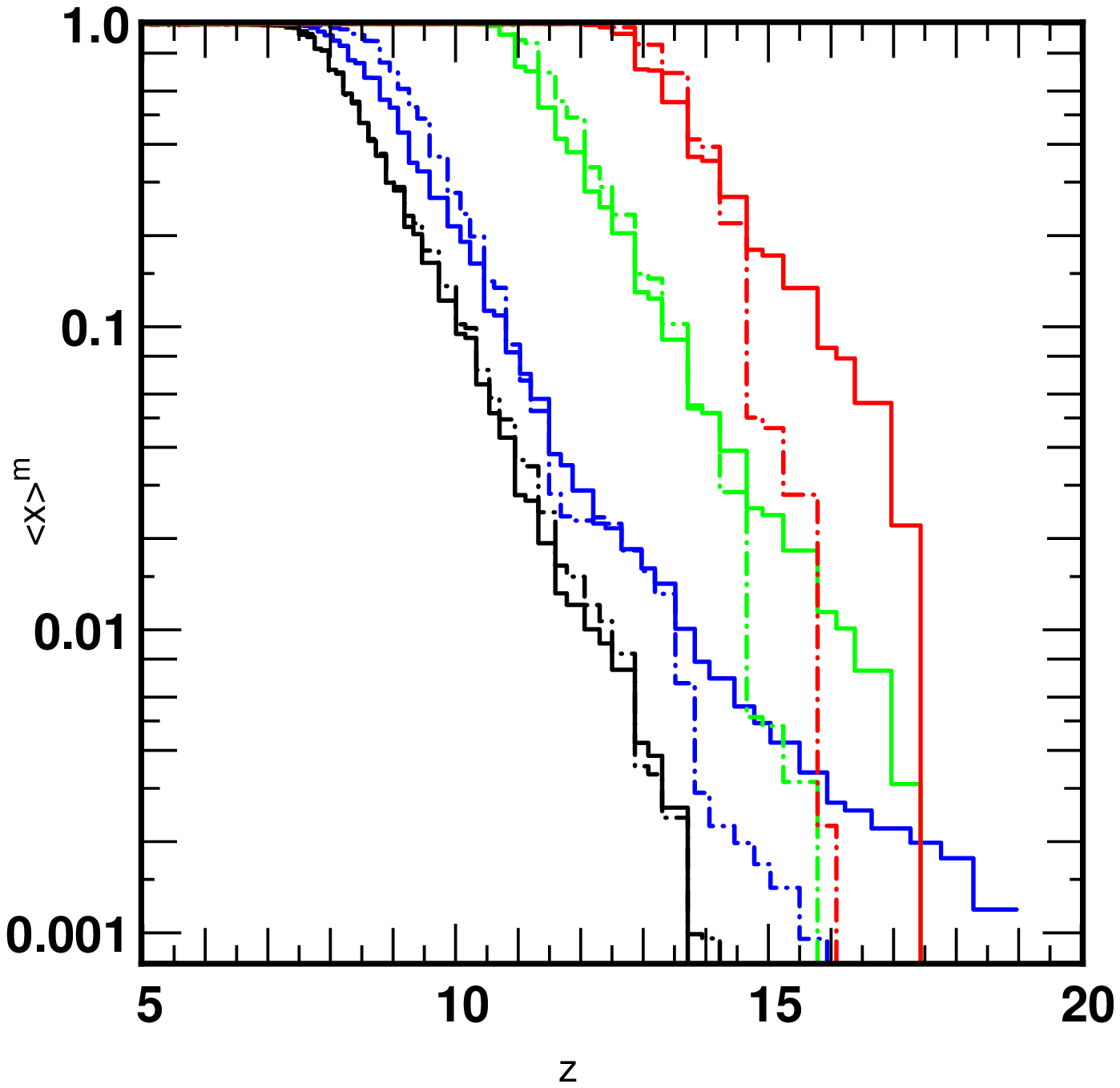}}\\
{\includegraphics[width=0.95\linewidth,clip]{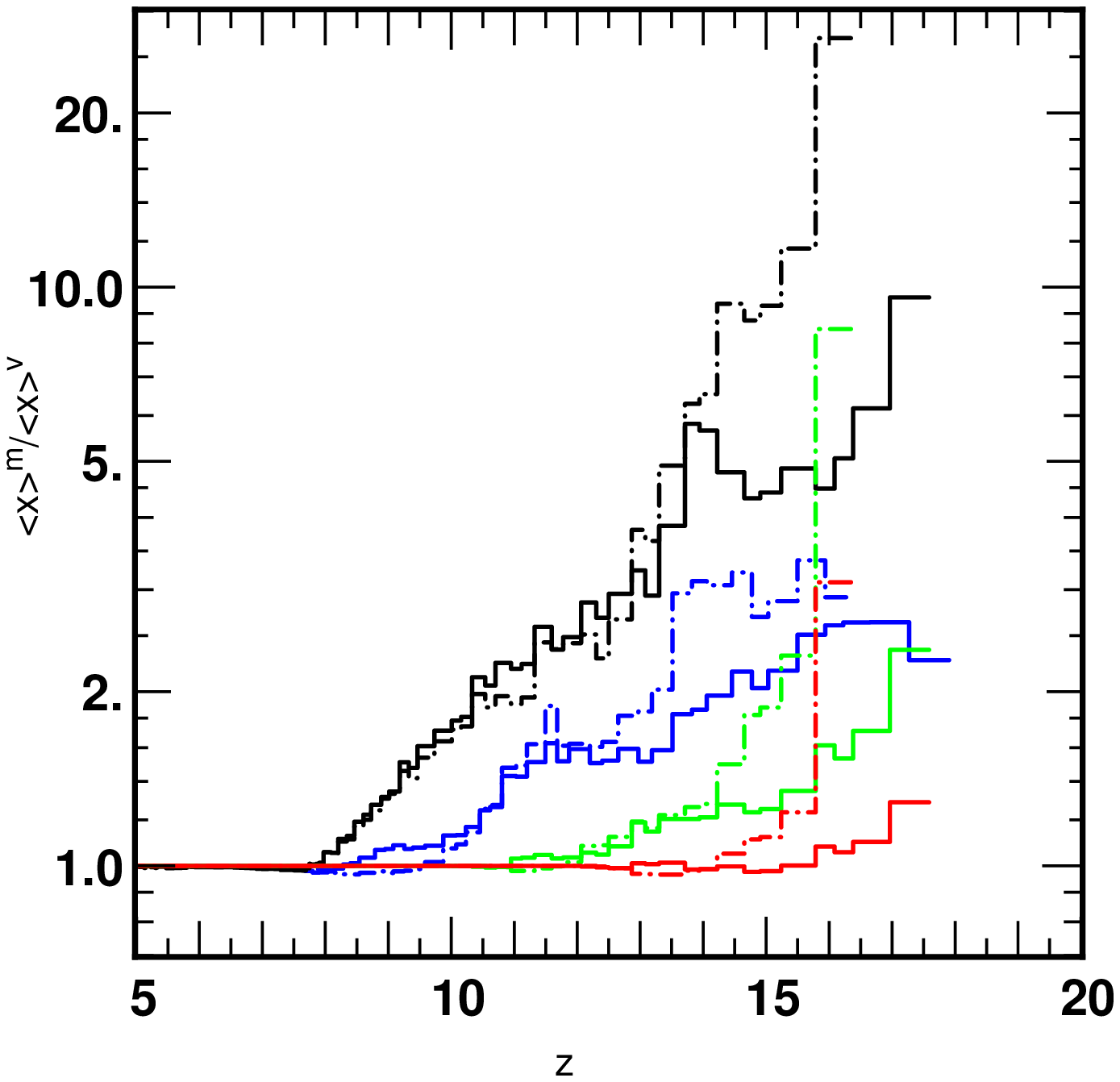}}
%\end{tabular}
\caption{Ionized fractions of the MW (solid line) and M31 (dot-dashed line) progenitors for our 4 baseline models. Top: mass-weighted, middle: log scale, bottom: ratio of the mass-weighted to the volume-weighted ionized fractions  $\xm/\xv$. The color and symbol code is specified in the top panel.}
\label{f:xs}
\end{figure}

%%%%%%%%%%%%%%%%%%%%%%%%%%%%%%%%%%%%%%%%%
\section{Results}
%%%%%%%%%%%%%%%%%%%%%%%%%%%%%%%%%%%%%%%%%
\label{s:results}

First of all, we check the good behaviour of our setup and method by analysing the global reionization history of the 2 major galaxies' progenitors in the box. Then we produce reionization maps to investigate how reionization proceeded spatially within the progenitors for our sets of models.

\subsection{Global progenitor reionization history}

The dark matter particles which end up within 300 $\hmo$ kpc of the MW center at z=0 are located in a sphere of $\sim 1$ \hmpc comoving radius at z=19.35. Their detailed distribution is not exactly spherical but the 1 \hmpc sphere remains a good approximation of it. This defines our MW progenitor volume, and we proceed similarly for M31. The evolution of the mass-weighted ionized fraction within the progenitor is shown in Fig. \ref{f:xs} for our H42-44 and SPH baseline models.
Depending on source emissivity, the reionization of the progenitors is achieved between z=7.5 and z=14. The difference in reionization histories due to the different emissivities is always much larger than the difference between MW and M31 reionization histories.
All models produce a very smooth and monotonous reionization of the progenitors.
The timings, and trends with emissivity are in fair agreement with a number of recent studies, such as \cite{li2013}.
The bottom panel of Fig. \ref{f:xs} shows the ratio of the mass-weighted to the volume-weighted ionized fraction. For the SPH, H42 and H43 models, $\xm/\xv >1$ for most of the redshift range, indicating that high-density regions are more ionized than low-density regions, i.e. I-front propagation proceeds locally inside-out: high density regions containing the sources are reionized first and the neighbouring regions with lower densities are impacted only later on. The ratio drops below unity as voids become dominant in ionized regions before reaching unity as sourceless denser regions such as gas filaments \citep{finlator2009} are eventually reionized. The maximum of $\xm/\xv$ (i.e. early times) drops with increasing emissivity, because in the photon-poor scenarios, it takes longer for the cell hosting the source to get ionized and let the photons leak out into the low-density regions.

\subsection{Reionization maps}

\begin{figure*}
\begin{tabular}{cc}
  {\includegraphics[width=0.49\linewidth,clip]{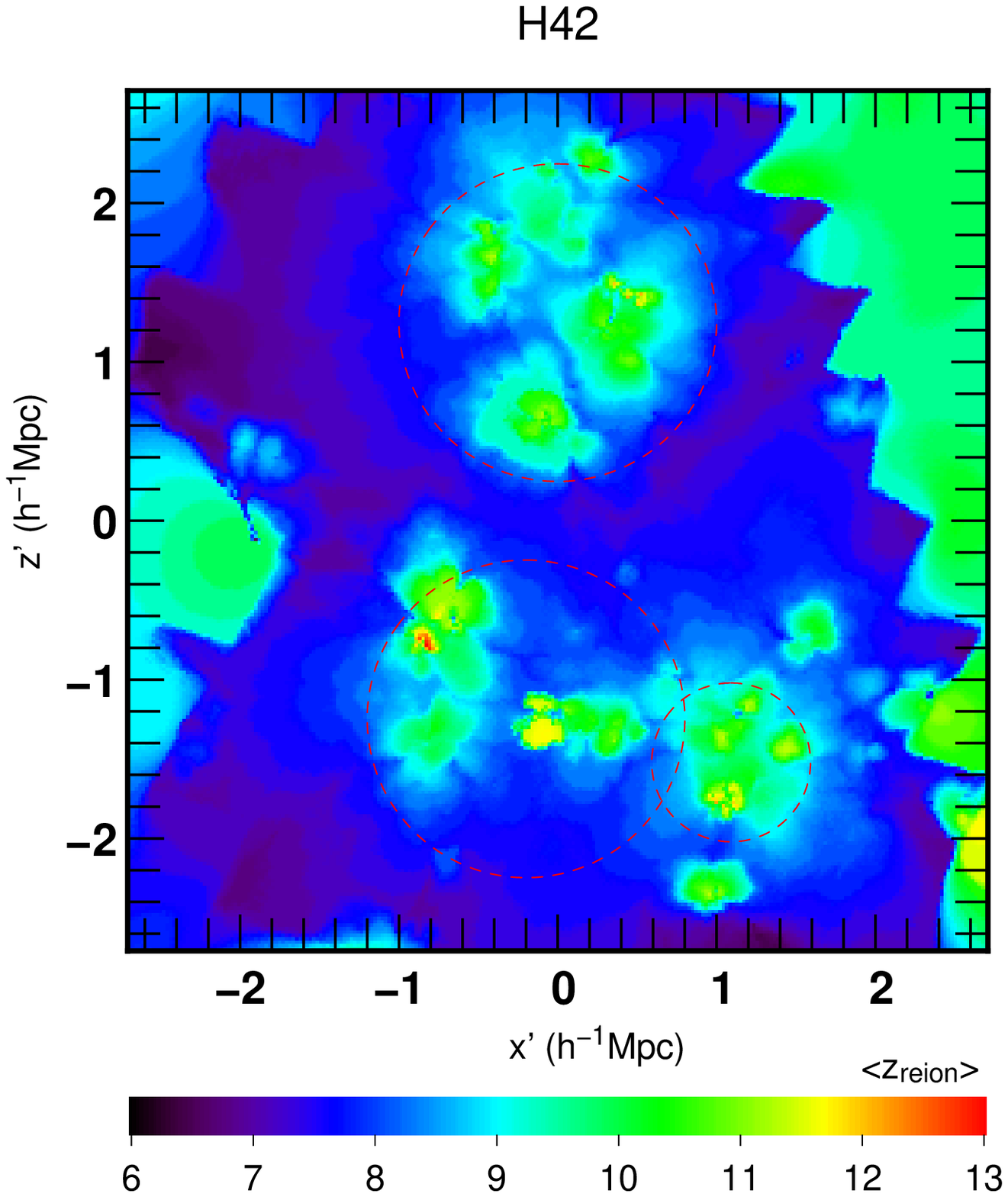}}&
  {\includegraphics[width=0.49\linewidth,clip]{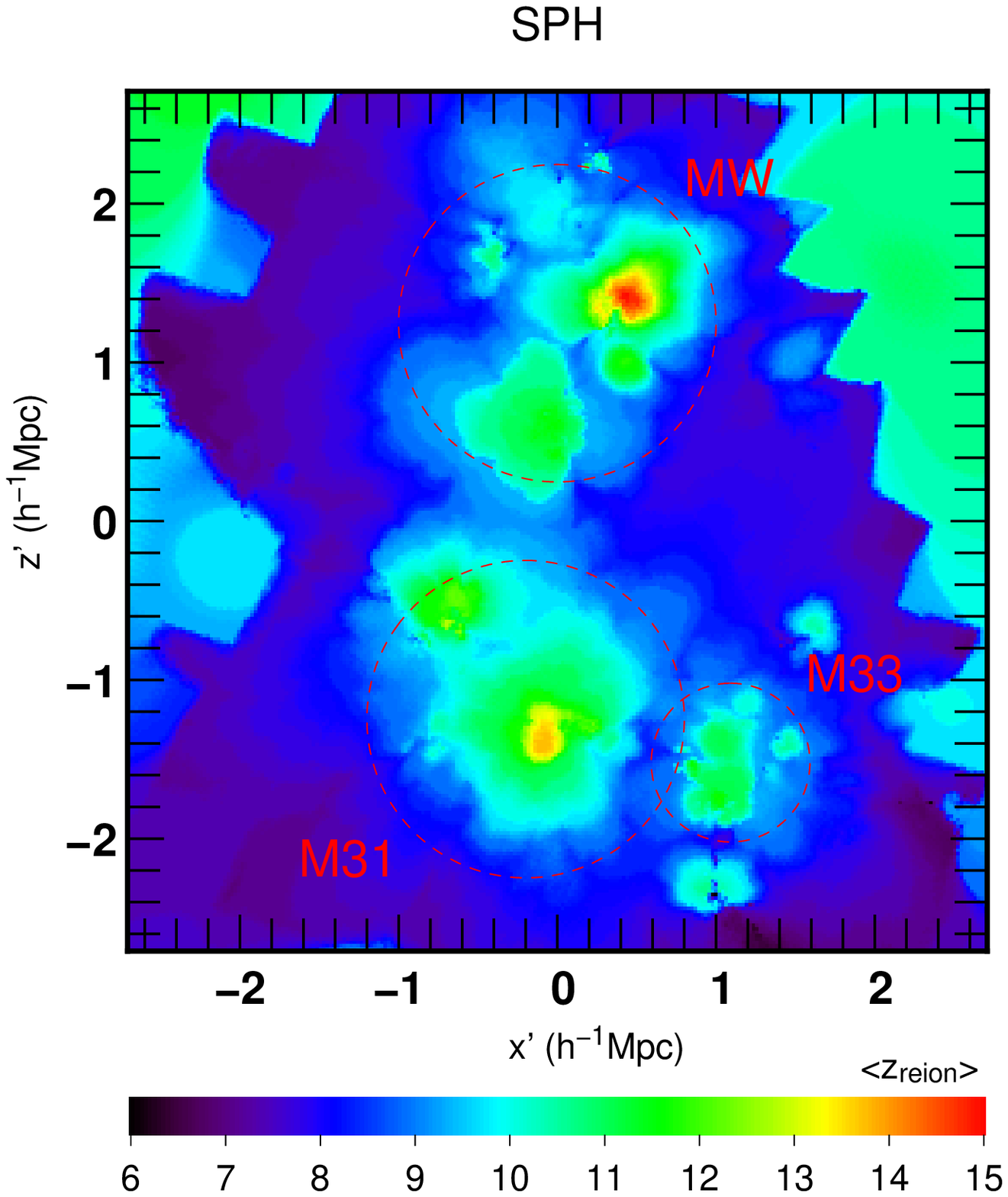}}\\
  {\includegraphics[width=0.49\linewidth,clip]{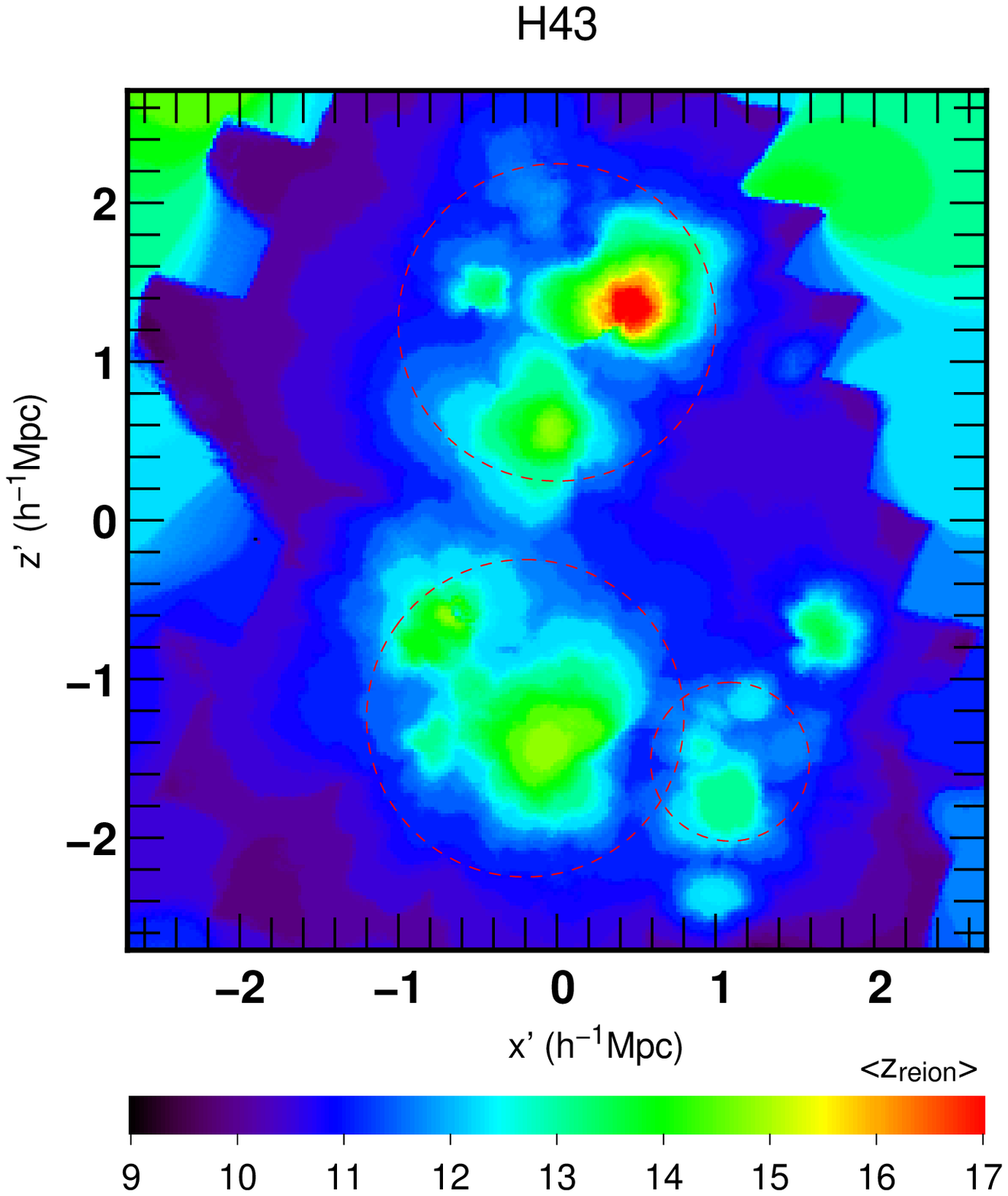}}&
  {\includegraphics[width=0.49\linewidth,clip]{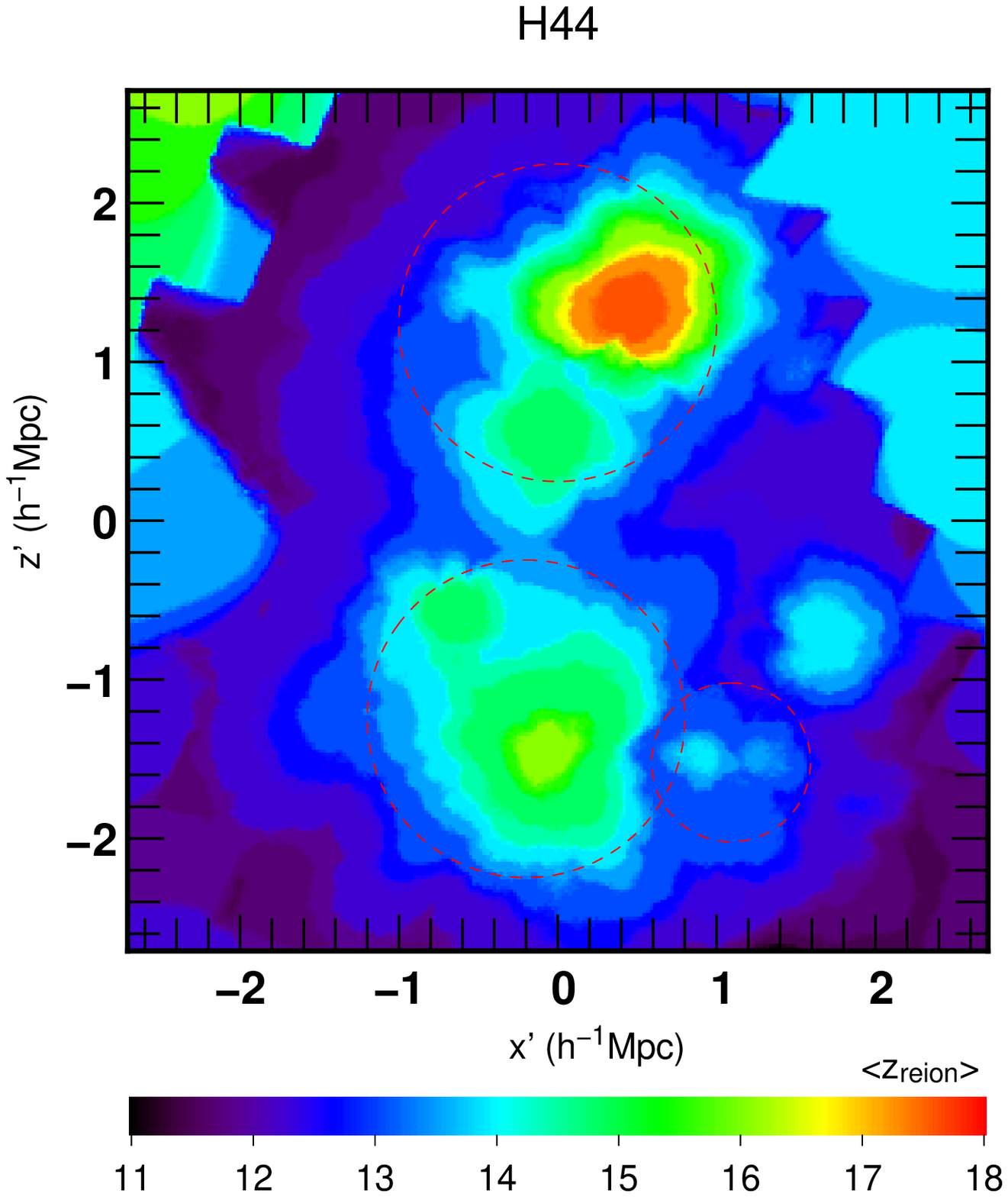}}
\end{tabular}
  \caption{Reionization maps of the local group of galaxies for our 4 baseline models. The simulation domain has been cut in the plane containing the centers of M31, M33 and the MW. The maps are obtained as the average $\zreion$ of a slab of 0.2 \hmpc thickness centered on this plane. The color codes the reionization redshift of each cell. The red dashed circles show the approximate location of the progenitors of the 3 major local group galaxies. The square artifacts in the corners are due to the transition from the high to low resolution domains of the SPH simulation.}
\label{f:rmaps}
\end{figure*}

%\subsection{Reionization maps}

We determine the reionization redshift of a given cell as the redshift of the last snapshot where its neutral fraction was above $\xion > 0.5$, i.e. if a cell reionizes at $z_1$ and then recombines, and gets ionized again at $z_2 < z_1$, we keep $z_2$ as the reionization redshift. 
The respective centers of the MW, M31 and M33 at z=6 progenitors define a unique plane, which we will refer to as the 3M plane throughout the paper. It provides a useful common reference for studying the reionization history of these galaxies.
The maps of Fig. \ref{f:rmaps} show the average reionization redshift in a slab of 0.2 \hmpc thickness centered on this plane. The 3 red dashed circles of  radius 1, 1, and 0.5 \hmpc respectively for MW, M31 and M33 are centered on each galaxy's center of mass at z=6.
They are indicative of the position and extent of each galaxy's progenitor. 
The sawtooth features on the sides are due to the transition from the HR to LR domain of the original hydrodynamical simulation. The maps are ordered in increasing global average reionization redshift from which also corresponds to increasing source efficiency. The color codes are set by the min and max redshifts of each map. This allows us to investigate the difference in I-front structures and propagation between the different models.

\subsubsection{Impact of source emissivity}
Low efficiency (H42 and SPH) models are patchier than their high efficiency counterparts (H43-44). 
The MW progenitor of the low efficiency models seems to consist of 4 main objects which reionize in isolation. This patchiness is expected due to the multiplicity of sources within each progenitor. The patches themselves also display a lot of internal structure. The latter can be related to the presence of dense infalling gas sheets and filaments, which reionize later than diffuse regions, as shown in \cite{finlator2009}. The structures of the individual patches also become smoother with increasing source efficiency.
%We also note that higher efficiency models yield earlier reionization, as expected. 
%The progenitors are fully reionized at z=6-7 in lowest efficiency models, and at z=12-13 in the most efficient model.
Due to the very low density we set in the LR region, reionization happens there very quickly, driven by a few sources outside of the volume plotted here, located at the HR/LR boundary.

\subsubsection{Star vs halo source model}
The H42 and SPH models have comparable emissivities at all times. As a result, The overall structure of their maps is rather similar: the patches are in comparable numbers and extents, except for the largest MW patch, which seems to reionize earlier in the SPH model.
However, the average $\zreion$ is slightly higher in the SPH model. Indeed, we have shown that the H42 model is slightly fainter, and its photons are distributed upon a larger number of sources, and therefore each source is less luminous than its SPH counterpart. 
This small difference in luminosity is also the cause of the difference seen in the M31 region between the H42 and SPH models. Indeed, for H43 and H44, it looks much more like the SPH.
Therefore, besides the effect described above, which could be compensated by a small increase in emissivity of the H42 model, we see that the details of source modelling play little role: photon output is the driving parameter here.

\subsubsection{Comparison with M31 and M33}
Except for the H42 model, the progenitor of M31 consists of less reionization patches than the MW region.
However, at high emissivities (H44), both objects only have 2 major patches.
%Moreover, MW's reionization seems to start earlier, as the highest $\zreion$ patch is always larger in the MW than in M31.
In the SPH model, reionization starts slightly earlier in the MW than in M31. Indeed, in the simulation we used, the first stars appear in larger number in the MW region than in M31. This trend is also seen in the H43-44 models, indicating that its origin likely lies in the slower assembly of M31 in the simulation. Indeed such a delay in the formation of 2 galaxies of similar mass in a pair is not uncommon, as can be seen in Fig. 2 of \cite{jaime2011} for similar simulations.
%This must be regarded as fortuitous rather than a general feature of local group formation.

\subsection{Reionization in isolation?}
In all our baseline models (H42,43,44) and SPH, M31 and the MW ignore each other during the reionization epoch. There is always a clear gap in $\zreion$ between them, and therefore the reionization of each progenitor is driven by its inner sources only. This is also true for the less massive M33, which seems to reionize a fair fraction of its progenitor in all models.
% although the color scale makes it difficult to see in panel (c). However, we can not say at this stage if some fraction of M33's progenitor is not reionized by photons from M31. 

The fact that MW and M31 did not interact radiatively during reionization could be an important simplification in satellite population models where no influence from Virgo is considered. This result would validate the approach used for the internal reionization models of \cite{ocvirk2011}, \cite{font2011} and many more, who neglect the influence of M31 on the reionization of MW satellites.

%\subsection{The strong feedback case}
\begin{figure*}[t]
  {\includegraphics[width=0.49\linewidth,clip]{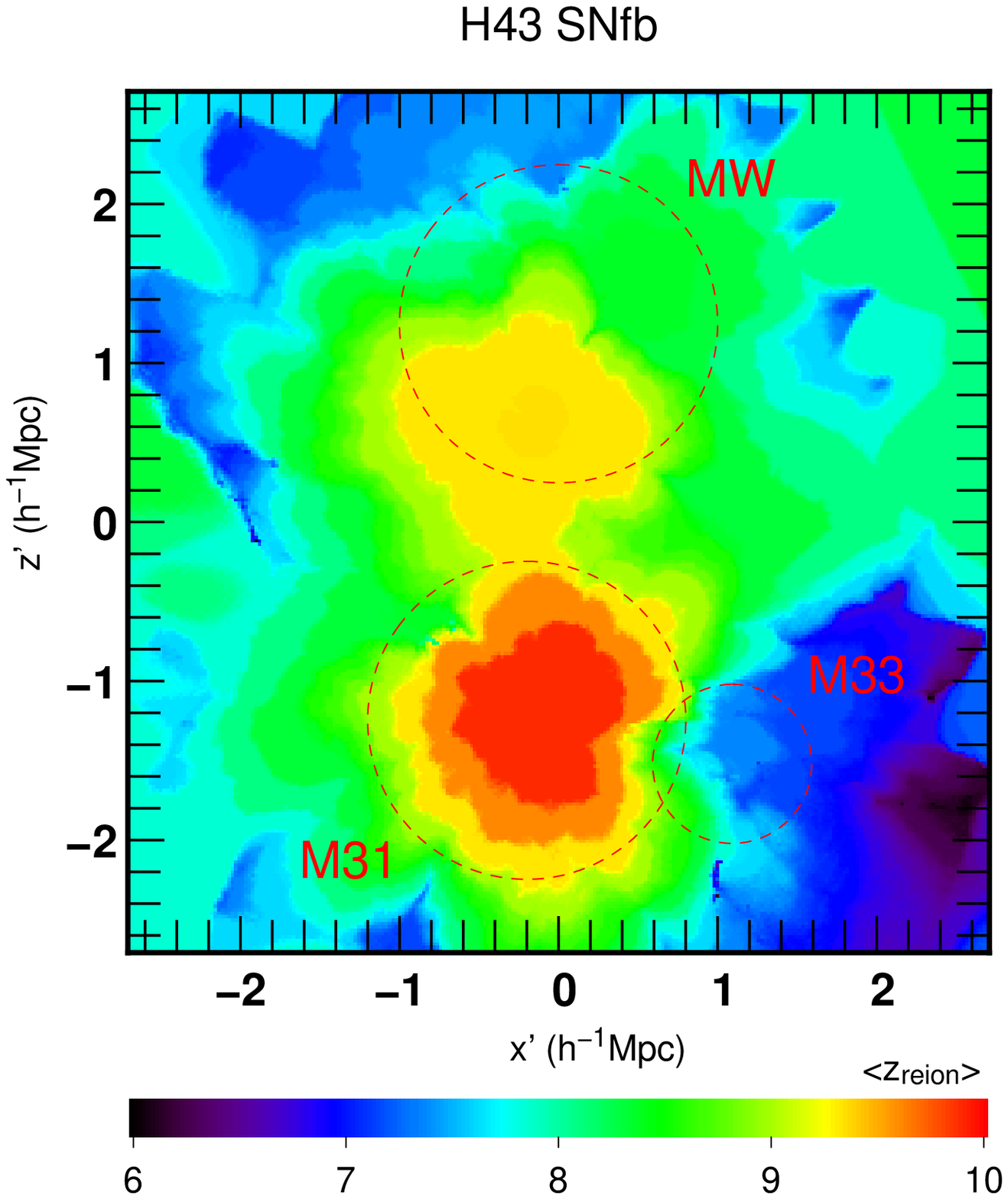}}
  {\includegraphics[width=0.49\linewidth,clip]{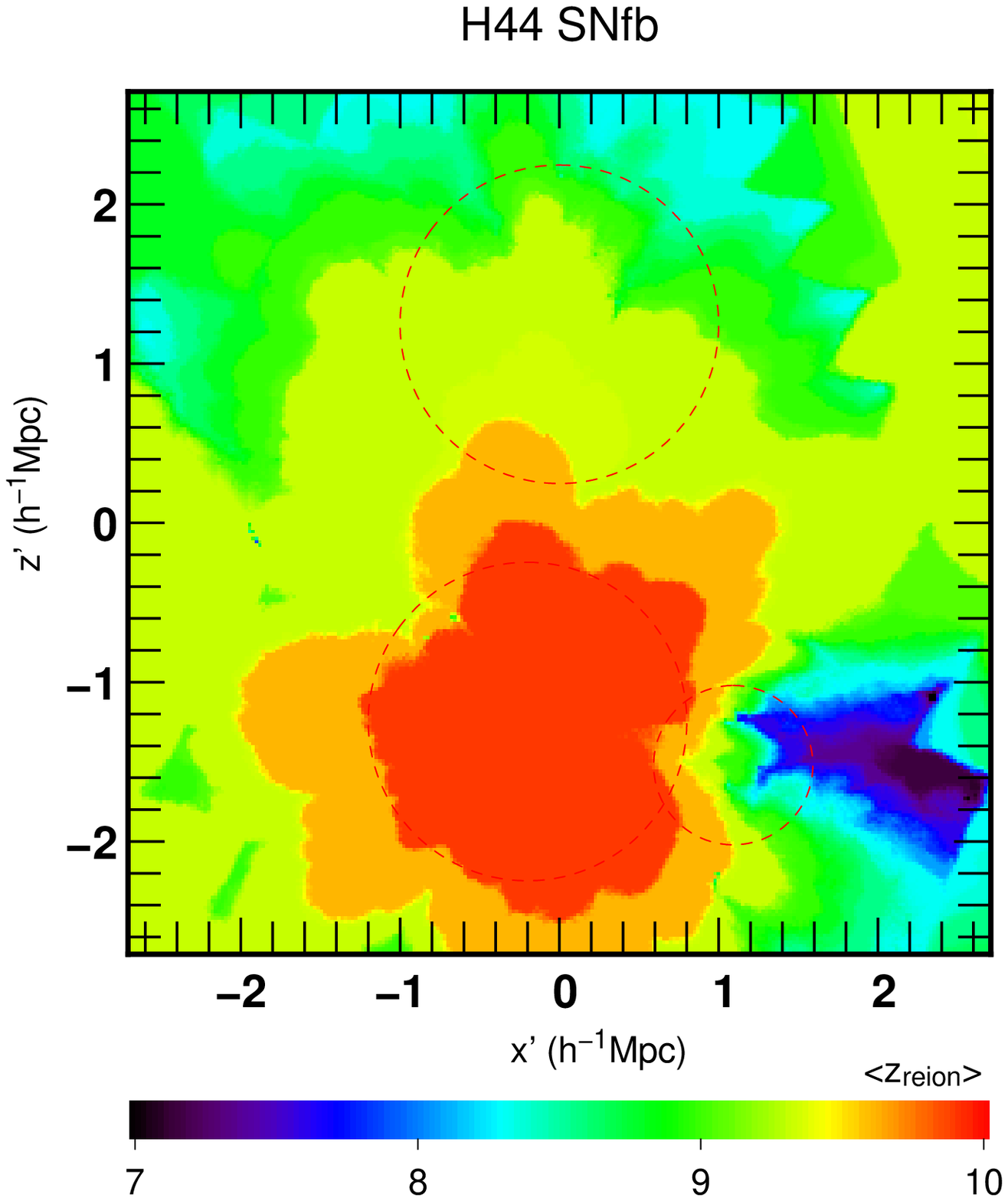}}
  \caption{Same as Fig. \ref{f:rmaps} for our 2 strong feedback scenarios. Reionization becomes very quick and is driven by a small number of sources, hence the small number of patches. In the most extreme scenario (right), M31 reionizes the MW progenitor.}
\label{f:rmapsSFB}
\end{figure*}

However, in the following, we show that there is still a possibility that the MW and M31 may have influenced each other during the EoR, by analysing our models with strong supernova feedback.
The global properties of these scenarios are shown in Tab. \ref{t:models}. Since the mass threshold for continuous star formation delays the apparition of the first efficient sources, the SNfb models reionize much later than the baseline models with the same emissivity. 
Model H42 SNfb can be readily dismissed as unrealistic since the reionization of the MW or M31 is achieved later than z=6. This model is at odds with the constraints derived from the Gunn-Peterson trough in quasar spectra \citep{fan2006}. Moreover, a broad range of reionization models agree in predicting that the MW or M31 should have reionized earlier than the rest of the Universe \citep{li2013,Chardin2013}. Therefore we consider only the 2 high emissivity H43 and H44 SNfb models, and compute their reionization maps, as shown in Fig. \ref{f:rmapsSFB}.
First of all, the maps have even less structure than in the H44 baseline model. Instead of 2 patches we have only one major patch in M31 and MW's progenitors in H43 SNfb and just one patch in the whole map for H44 SNfb, centered on M31.
While M31's progenitor displays a clear inside-out reionization pattern in both maps, the situation is more complex for the MW's progenitor. In the H43 SNfb model, the MW region features a single patch, produced by a source which turned on much later than M31's sources. Moreover, a significant fraction of the progenitor's volume appears to reionize at the same redshift as the IGM surrounding the MW and M31. Whether the progenitor is internally or externally reionized is not so clear-cut anymore. 
The situation becomes unambiguous in the H44 SNfb panel: the map shows one single patch centered on M31. In this scenario, the reionization of the LG is driven by M31 alone. The MW's progenitor is quickly, externally reionized by M31.
%One should keep in mind, though, that the mass threshold used to account for supernova feedback is quite high: 

%Here we show that when there is a very strong internal feedback preventing low mass sources from being efficient emitters, combined with high emissivity, galaxies in pairs can start to influence each other.

%The first sources to appear reionizes a large volume. Moreover, since they appear later, they are pouring their photons into a lower density IGM than their high-z counterparts because of universal expansion.

%%%%%%%%%%%%%%%%%%%%%%%%%%%%%%%%%%%%%%%%%%%%%
\section{Discussion}
%%%%%%%%%%%%%%%%%%%%%%%%%%%%%%%%%%%%%%%%%%%%%
\label{s:discussion}

%While previous studies such as \cite{weinmann2007,li2013} focused on the radiative influence of Virgo on the LG, here we investigated the interplay between the MW and M31 progenitors during the EoR. 
Even though most of our models yield clearly isolated reionization scenarios for MW and M31, a radiative influence of M31 on MW or vice versa can not be ruled out with our current constraints on the strength of the feedback and the emissivity of the sources. Reducing the number of sources by increasing the minimum mass threshold while increasing their emissivity  yields a ``first source wins it all'' scenario. We note however that this scenario is at the extreme end of our parameter range: it appears only with the strongest source suppression and highest emissivity. It is not clear if this model would produce a reasonable global reionization history in a large scale box. It would be useful to check this with a further simulation of a large volume, but this is outside of the scope of the present paper. Reviewing the literature, we did not find any study where this combination of emissivity and source minimum mass has been ruled out.

The internal, isolated reionization of MW-size galaxies found by our baseline models (H42,43,44 and SPH) is in agreement with results from \cite{weinmann2007} and the photon-poor regime of I11. However, the model 1 of I11 (roughly similar to our model H43) predicts that for high emissivities the LG should be reionized {\em externally} by photons from Virgo, at z=10.5-10.25. We found that in this regime the MW and M31 actually reionize much earlier than this, i.e. at z=11.5-11.8, and they do so {\em internally}. 
This highlights an important caveat of the present study: since the Virgo galaxy cluster is outside of the HR region of the CLUES simulation we used, it is simply not taken into account. Moreover, our work and I11 use different setups, which makes comparing reionization timings very tricky.
%Therefore one could be drawn to the conclusion that, contrary to the predictions of I11, the reionization of the local group is always internally driven, within the reasonable, yet wide range of emissivities we consider, and despite the presence of the nearby Virgo cluster. However, eventhough we use similar emissivities as I11, our study can not be compared directly.
%\begin{itemize}
%\item{first of all we do not use the same simulation, and therefore the mass assembly history of the galaxies is likely to differ. To check this, the experiment presented here should be performed on more high resolution realizations of the formation of the LG, in order to check that our result is not just due a fortuitous quick formation of the MW. This is not likely though, since M31 achieves full reionization at about the same epoch as the MW, eventhough it starts slightly later.}
%First, unlike I11, we have not implemented any explicit feedback on the low-mass, suppressible sources. It is difficult to predict the impact of this simplification: although we do not suppress low-mass sources in the H43 model, they are 15 times fainter than in model 1 of I11.
Furthermore, the mass resolution of our simulation is a factor $\sim 50$ higher than in I11. It is therefore expected that the lowest mass haloes form earlier in our simulation, providing earlier sources to reionize the LG.
This is well documented in the literature. 
For instance, the top panel of Fig. 1 in \cite{aubert2010} shows the total number of emitted photons per hydrogen atom for 4 different resolutions, with the same stellar emissivity for all 4 simulations. At z$>10$, the 12.5 \hmpc box can produce up to 100 times more photons than the 50 \hmpc box. This corresponds to a factor 64 increase in mass resolution. This is close to the difference in mass resolution between I11 and the present study. We note however that \cite{aubert2010} refers to a postprocessing based on using star particles as source, and not a halo-based formalism as we do here and in I11. Furthermore, our crude implementation of feedback is based on a filtering of haloes less massive than $10^7 - 10^8 \Msun$ (HXX models) and $10^9 \Msun$ (HXX SNfb models), which reduces the impact of increased resolution on the number of sources.
Finally, considering this result, it seems reasonable to expect that z$>10$ photon production can be boosted by a large factor as a result of improved mass resolution between I11 and our study. This boost in turn leads to an earlier reionization of the LG, despite using emissivities roughly similar to the photon-rich scenario of I11. Likewise, increasing resolution would also boost the photon output of Virgo at early times, so that its I-front could reach the LG earlier than found by I11. But will it happen early enough to reionize the LG externally rather than internally? The question remains.
Ideally, assessing the influence of Virgo on the reionization of the LG will require a high resolution simulation of the formation of the LG and its environment, including Virgo, in the spirit of I11, but at the resolution of the present work or better. 
Thanks to the rapid evolution of the hardware and high performance computing facilities, we expect this will become feasible in the coming years.

%\pier{Note que les predictions de Alvarez 2009 sont aussi probablement pourries a cause de sa resolution. A basse resolutino on a surement des effets de reionisatino externe spurieuse, alors que si on avait tout en HR on verrait que meme les petits objets voisins de gros reionisent interne}

%\pier{There is also the possiblity of a mixed origin for MW reionization:
%the innermost regions of the MW may have been fully reionized internally while the outskirts of the progenitor would have been reionized by both MW sources and by photons from Virgo.}

%%%%%%%%%%%%%%%%%%%%%%%%%%%%%%%%%%%%%%%%%
\section{Conclusions}
%%%%%%%%%%%%%%%%%%%%%%%%%%%%%%%%%%%%%%%%%
\label{s:conclusions}
We have radiatively post-processed a high resolution simulation of the formation of the local group in order to investigate the reionization of MW and M31 at galactic scale. We have used 7 different ionizing source models with various emissivities and star formation criteria to assess the impact of uncertainties on the source properties.
When considering only atomic cooling haloes as sources, we find that the reionization of the MW progenitor is generally patchy, with 2-4 major regions and a few minor ones reionizing in isolation. Increasing emissivity leads to fewer isolated patches and accelerated reionization: our most photon-poor scenario reionizes the MW progenitor at z=8.4, and at z=13.7 in the most photon-rich regime.
Our results are in fair agreement with the literature available, although very few studies tackled reionization at these scales. 
%Suppressing star formation in haloes less massive than $10^{9} {\rm \Msun}$ , for instance due to strong supernova feedback, leads to a delayed reionization with respect to our baseline models, with less structure, fewer patches within the progenitors.
In all models except the most extreme, the MW and M31 progenitors reionize in isolation, despite being relatively close to each other even during the EoR. The corresponding reionization maps always show a clear gap in $\zreion$ between the two progenitors. Only in the case of strong supernova feedback suppressing star formation in haloes less massive than $10^{9} {\rm \Msun}$, and using the highest emissivities, we find that the MW is reionized by M31, which hosts the first efficient source to appear in the box.

Putting our study back into the general context of galaxy formation, this work is an additional step in the investigation of the internal versus external nature of galactic reionization for the Milky Way and M31, as a function of feedback type and source emissivity. Further effort should pe put into exploring the source parameter space, along with improving feedback models, radiative and hydrodynamical, extending this work to a larger number of realizations of the MW-M31 system, and, in particular, including the impact of Virgo.
We hope this work will help build a more accurate and sensible framework for future semi-analytical models of galaxy formation during the EoR, as well as models of the satellite population of the MW and M31.

\section*{Acknowledgements}

We thank the anonymous referee for providing constructive comments, which helped to significantly improve our models and the paper. This study was started in the context of the LIDAU project\footnote{\url{http://aramis.obspm.fr/LIDAU/Site_2/LIDAU_-_Welcome.html}}. The LIDAU project was financed by a French ANR (Agence Nationale de la Recherche) funding (ANR-09-BLAN-0030). The RT computations were performed using HPC resources from GENCI-[CINES/IDRIS] (Grant 2011-[x2011046667]), on the hybrid queue of titane at Centre de Calcul Recherche et Technologie, as well as Curie, during a grand challenge time allocation (project PICON: Photo-Ionisation of CONstrained realizations of the local group). The CLUES simulations were performed at the Leibniz Rechenzentrum Munich (LRZ) and at the Barcelona Supercomputing Center (BSC). AK is supported by the {\it Spanish Ministerio de Ciencia e Innovaci\'on} (MICINN) in Spain through the Ram\'{o}n y Cajal programme as well as the grants CSD2009-00064 and CAM~S2009/ESP-1496 and the {\it Ministerio de Econom\'ia y Competitividad} (MINECO) through grant AYA2012-31101. He further thanks Peter Thomas for angles who burn their wings. GY acknowledges support from MICINN under research grants AYA2009-13875-C03-02, FPA2009-08958 and Consolider Ingenio SyeC CSD2007-0050. The author thanks C. Scannapieco for precious hints dispensed in the initial phase of the project, as well as the CLUES collaborators for useful discussions. The author thanks D.~Munro for freely distributing his Yorick programming language\footnote{\url{http://www.maumae.net/yorick/doc/index.html}}, and its yorick-gl extension.

\bibliographystyle{apj}
\bibliography{mybib}
%%%%%%%%%%%%%%%%%%%%%%%%%%%%%%%%%%%%%%%%%%%%%%%%%%%%%%%%%%%%%%%%%%%%%%%%%%%%%%%
%

\label{lastpage}
\end{document}